\documentclass[aps,prb,10pt,twocolumn,superscriptaddress,floatfix]{revtex4-2}
\usepackage[utf8]{inputenc}
\usepackage{tabularx}
\usepackage{array}
\usepackage{float}
\usepackage{graphicx}
\usepackage{longtable}
\usepackage{amsmath}
\usepackage{amssymb}
\usepackage{textcomp}
\usepackage[dvipsnames]{xcolor}
\usepackage{gensymb}
\usepackage{titlesec}

\definecolor{map}{rgb}{108, 0, 108}

\definecolor{pink}{rgb}{1, 0.5, 0}

\definecolor{darkgreen}{rgb}{0, 0.5, 0}

\begin{document}

\title{Incommensurate and commensurate antiferromagnetic orders in the kagome compound UV$_{6}$Sn$_{6}$}
\author{Midori Amano Patino}
\affiliation{Univ.~Grenoble Alpes, CNRS, Grenoble-INP, Institut N\'eel, 38000 Grenoble, France}
\affiliation{Univ.~Grenoble Alpes, CEA, Grenoble-INP, IRIG, Pheliqs, 38000 Grenoble, France}
\author{Stephane Raymond}
\affiliation{Univ.~Grenoble Alpes, CEA, IRIG, MEM, MDN, 38000 Grenoble, France}
\author{Georg Knebel}
\affiliation{Univ.~Grenoble Alpes, CEA, Grenoble-INP, IRIG, Pheliqs, 38000 Grenoble, France}
\author{Pierre Le Berre}
\affiliation{Univ.~Grenoble Alpes, CEA, Grenoble-INP, IRIG, Pheliqs, 38000 Grenoble, France}
\author{Stanislav Savvin}
\affiliation{Institut Laue Langevin, Grenoble, France}
\author{Elise Pachoud}
\affiliation{Univ.~Grenoble Alpes, CNRS, Grenoble-INP, Institut N\'eel, 38000 Grenoble, France}
\author{Eric Ressouche}
\affiliation{Univ.~Grenoble Alpes, CEA, IRIG, MEM, MDN, 38000 Grenoble, France}
\author{James C. Fettinger}
\affiliation{Department of Chemistry, University of California, Davis, California 95616, USA}
\author{Olivier Leynaud}
\affiliation{Univ.~Grenoble Alpes, CNRS, Grenoble-INP, Institut N\'eel, 38000 Grenoble, France}
\author{Jacques Pecaut}
\affiliation{Univ.~Grenoble Alpes, CEA, IRIG-SyMMES, 38000 Grenoble, France}
\author{Peter Klavins}
\affiliation{Department of Physics and Astronomy, University of California, Davis, California 95616, USA}
\author{Klaus Hasselbach}
\affiliation{Univ.~Grenoble Alpes, CNRS, Grenoble-INP, Institut N\'eel, 38000 Grenoble, France}
\author{Jean-Pascal Brison}
\affiliation{Univ.~Grenoble Alpes, CEA, Grenoble-INP, IRIG, Pheliqs, 38000 Grenoble, France}
\author{Gerard Lapertot}
\affiliation{Univ.~Grenoble Alpes, CEA, Grenoble-INP, IRIG, Pheliqs, 38000 Grenoble, France}
\author{Valentin Taufour}
\email{vtaufour@ucdavis.edu}
\affiliation{Department of Physics and Astronomy, University of California, Davis, California 95616, USA}

\begin{abstract}
We report on the synthesis of single crystals of the kagome compound, UV$_6$Sn$_6$, and present the results of magnetization, electrical resistivity, heat capacity, x-ray, and neutron diffraction experiments to characterize the structure and magnetic properties. UV$_6$Sn$_6$ crystallizes in a large supercell of the HfFe$_6$Ge$_6$ parent structure with an hexagonal symmetry in which some of the U atoms are shifted by $c/2$ in an ordered fashion. Below $T_{N1}\approx29$\,K, an incommensurate magnetic structure with a temperature-dependent wave vector $(0,0,k_z)$ is observed. Below $T_{N2}=23.5$\,K, the wave vector locks in to $(0,0,0.5)$, forming an antiferromagnetic ground state. The U moments align along the \(c\) axis retaining a large magnetic anisotropy. These findings highlight the role of the $5f$ orbitals from uranium in this structural family in driving both magnetic ordering and structural modulation and distinguish UV$_6$Sn$_6$ from its lanthanide-based analogs.
\end{abstract}

\maketitle

\section{Introduction}

Materials containing kagome networks have attracted a lot of attention for the study of topologically nontrivial phases, and intertwined orders~\cite{Yin2022Nature,Xu2023RPP,Wang2023NRP}. Among them, the 166 family of compounds, characterized by the general formula \(RT_6X_6\), represents a fertile platform for investigating the rich interplay of topology, electron correlations, and magnetism. The \textit{R} element can be a rare-earth element or Li, Mg, Ca, Ti, Zr, Hf, T is a transition metal element, and \textit{X} is typically a tetrel element such as Sn or Ge. In these materials, the kagome lattice, a two-dimensional network of corner-sharing triangles, facilitates the emergence of exotic phenomena, including charge density waves (CDWs)~\cite{Arachchige2022PRL,Pokharel2023PRM,lee_nature_2024}, phonon softening~\cite{Korshunov2023NC,Hu2024NC166,Fu2025PRB}, gapped Dirac fermions and surface states~\cite{Yin2020Nature,Pokharel2021PRB,Peng2021PRL}, a nodal ring~\cite{He2024PRB}, anomalous and topological Hall effects~\cite{Asaba2020PRB174415,Dhakal2021PRB,Wang2021PRB014416}, anomalous Nernst effect~\cite{Roychowdhury2022AM,Zhu2024PRB}, and an extraordinary structural diversity~\cite{Buchholz1981ZAAC,Venturini2006ZfK,Venturini2008CMA,Fredrickson2008JACS,Braun2023JSSC,Liu2023CPL,Yang2024CPB,Ortiz2025JACS}.

This combination of structural and electronic features has made the 166 family an important class of materials for studying the interplay of magnetic order, topology, and electron correlations. In this paper, we report on the synthesis and characterization of a recent member of the 166 family, UV$_6$Sn$_6$. The introduction of uranium in the lattice structure may give rise to rich magnetic and electronic properties due to the strong spin-orbit coupling and the duality between localized and itinerant character of the uranium 5$f$ electrons. We find evidence for a superstructure in the $ab$ plane in UV$_6$Sn$_6$, where additional Bragg peaks are observed both in x-ray and neutron diffraction experiments. Below $T_{\textrm{N}_1} \simeq 29$\,K, the U moments order antiferromagnetically in an incommensurate magnetic structure, as detected by neutron scattering. 
A commensurate antiferromagnetic ground state is observed below 23\,K. From resistivity and heat capacity data, we show that UV$_6$Sn$_6$ is a weakly correlated metal.


\section{Methods}

Single crystals of UV$_6$Sn$_6$ were grown using a self-flux method similar to that reported for YV$_6$Sn$_6$~\cite{Pokharel2021PRB}. Natural uranium ingots (99.27\% $^{238}$U, 0.72\% $^{235}$U, trace $^{234}$U, from an internal stock at CEA Grenoble) were cut in $\sim$0.5~g pieces and cleaned by electrolytic etching to remove the oxides on the surface~\cite{Taufour2011PhD}. Some crystals were grown at UC Davis following the same procedure with natural uranium from the NBL Program office (99.24\% $^{238}$U, 0.72\% $^{235}$U, trace $^{234}$U). Vanadium pieces (99.9\%, Aldrich) and tin shots (99.999\%, Thermo Fisher Scientific) were cleaned with 30\,vol.\% HCl (Sigma-Aldrich). The clean metals were introduced in a Canfield crucible set~\cite{Canfield2016PM,Canfield2025ZAAC} in the 1:6:20 molar ratio, and sealed in a silica ampoule with 1/3\,atm argon gas. The ampoule was heated to 1150\,\degree C in 7\,h, held there for 36\,h, slowly cooled to 1050\,\degree C in 287\,h. At 1050\,\degree C, the formed crystals were decanted by removing the ampoule from the furnace, inverting it, and immediately spinning it in a centrifuge~\cite{PaglioneButchRodriguez2021book}. Only crystals of UV$_6$Sn$_6$ were obtained with this temperature profile. The experiments with cooling to 780\,\degree C resulted in additional single crystals of USn$_3$. The as-synthesized UV$_6$Sn$_6$ single crystals exhibit a metallic gray color and hexagonal pyramidal morphology with edge lengths of 0.3–0.8\,mm (volume $\sim$0.02–0.33\,mm$^3$, Fig.~\ref{Fig:Structure}d). Laue x-ray diffraction confirms that the $c$ axis is oriented perpendicular to the hexagonal base, while the $a$ and $b$ axes extend toward the in-plane hexagonal vertices.

Single-crystal x-ray diffraction data were collected at 100–150 K using the four-circle Kappa geometry diffractometers: Rigaku XCallibur (Rigaku CrysAlis CCD software), and Bruker D8 Venture (APEX software). In both instruments, a \(\text{Mo } K_{\alpha}\) radiation source was used (0.71073~Å). The structural refinement was carried out using the crystallography software package GSAS-II~\cite{Toby2013JAppCrys}.

Magnetization measurements were performed using an MPMS3 SQUID magnetometer (Quantum Design) in VSM scan mode. The crystallographic axes of the sample were determined using Laue diffraction followed by polishing to obtain a flat surface perpendicular to the $c$ axis.

Resistivity measurements were performed in a physical property measurement system (PPMS, Quantum Design) using a four point lock-in technique; 15\,$\mu$m Au wires were spot welded to the sample for electrical contacts. Due to the irregular shape of the sample and the unperfect alignment of the electrical contacts, absolute values of the resistivity were not determined. The specific heat was measured using the PPMS with the heat pulse relaxation method; heat pulses of 0.5, 1, and 2\% of the actual temperature gave similar results. The samples for specific heat had masses of 1.45 and 0.9~mg, thus the specific heat of the sample was small compared to the addenda contribution.

Neutron diffraction experiments were carried out at Institut Laue Langevin, Grenoble on the two-axis diffractometer D23 and the large solid angle detector diffractometer XtremeD. A single-crystal sample with a mass of 1.48\,mg was pre-aligned using x-ray diffraction (Laue method) with the $b$ axis oriented vertically, so the horizontal scattering plane was ($a^*$, $c^*$) in a hexagonal setting. D23 is equipped with a lifting detector reaching 30\degree~and neutrons of wavelength 1.28\,{\AA} and 2.36\,{\AA} were used with a copper monochromator and a pyrolitic graphite (PG) monochromator, respectively. In this latter case, a PG filter was inserted in the incoming beam to remove higher order contamination. On XtremeD, the wavelength of 2.45\,{\AA} provided by a PG monochromator was used, the angular coverage of the detector was of 130\degree~in the horizontal plane and 24\degree~in the vertical plane. A PG filter was also installed. An oscillating radial collimator was placed in front of the detector bank to suppress the sample environment contribution to the collected data and reduce background. For both experiments, the sample was inserted in an helium flow cryostat. The MAG2POL software was used to refine the nuclear and magnetic structures \cite{Qureshi2019JAC}.

\section{Results and discussion}

\subsection{Crystal structure}
\begin{figure*}[!htb]
\centering
\includegraphics[width=\linewidth]{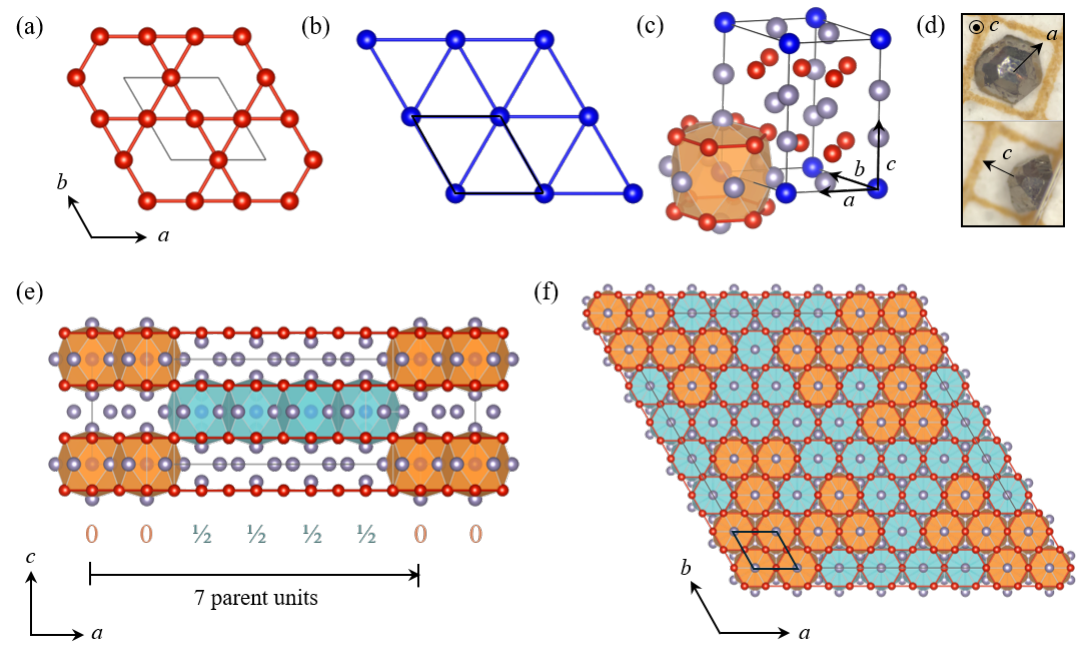}
\caption{\label{Fig:Structure} The HfFe$_6$Ge$_6$-type parent structure of UV$_6$Sn$_6$, where the V atoms (red) form a kagome network (a), and the U atoms (blue) form a triangular lattice (b). The V kagome and U triangular lattices arrange in layers to form the structure depicted in (c), where the V and Sn atoms around U form a hexagonal cavity represented as a polyhedron. A picture of a representative single-crystal sample on a millimeter grid is shown in (d). The ordering of the cavities can lead to patterns such as the one shown in (e). These patterns extend the structures of \(RT_6X_6\) compounds into supercells, like the one depicted in (f). The structure of UV$_6$Sn$_6$ can be described as a supercell where a periodic modulation of the HfFe$_6$Ge$_6$-type parent cell---marked in black in (f), yields a 7 × 7 superstructure on the \(ab\) plane. The hexagonal cavities occupied by a U shift along the \(c\) axis by \(c/2\) yielding the pattern shown in (e). This pattern is identical along the \(b\) crystallographic axis (\(a = b\)).}
\end{figure*}

The structure of UV$_6$Sn$_6$ can be described as a $7\times 7$ modulation of a HfFe$_6$Ge$_6$-type cell on the $ab$ plane. In the latter \textit{parent} cell (hexagonal symmetry, space group $P6/mmm$), the V atoms form a kagome network and the U atoms form a triangular lattice which stack in layers along the $c$ crystallographic axis (Fig.~\ref{Fig:Structure}a-c). Ignoring
the superstructure first, a model of this HfFe$_6$Ge$_6$-type parent cell can be refined against single-crystal x-ray diffraction data with full occupancy for all sites, as summarized in Table~\ref{Table:Ref_HfFe6Ge6-type} of Appendix~\ref{Appen:Details_HfFe6Ge6-type}.

Although the goodness-of-fit value of 1.76 (GOF, Table~\ref{Table:Ref_HfFe6Ge6-type} in Appendix~\ref{Appen:Details_HfFe6Ge6-type}) suggests reasonable agreement between the measured and calculated average intensities, the high value of the weighted profile factor, \(wR2\)~=~20.76\%, reveals a significant discrepancy between the observed and calculated structure factors.

\begin{figure*}[!htb]
\centering
\includegraphics[width=\linewidth]{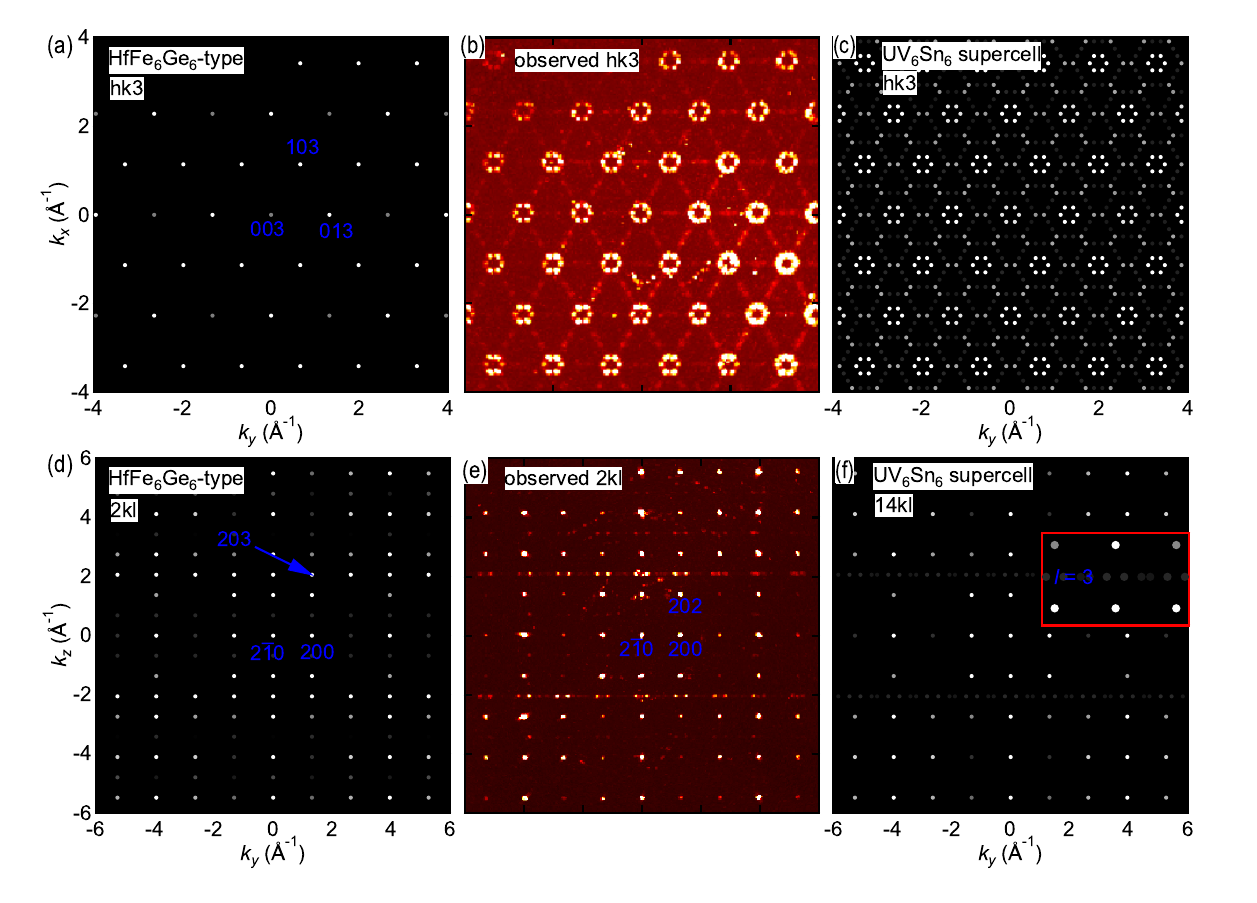}
\caption{\label{Fig:Precession_hk3-2kl} Comparison of the calculated and observed reconstructions of the reciprocal space centered on planes \(hk3\) and \(2kl\) for UV$_6$Sn$_6$. The calculated diffraction patterns in (a) and (d) show the expected Bragg spots (reflections, marked in blue) for a HfFe$_6$Ge$_6$-type cell on planes \(hk3\) and \(2kl\), respectively. The observed \(hk3\) pattern in (b) shows undetectable intensity at the expected Bragg spots and the presence of six satellite reflections around them. Additional satellite reflections are observed along \(h\) and \(k\) directions. The observed \(2kl\) pattern in (e), indexed in the HfFe$_6$Ge$_6$-type cell, shows discrepancies in the intensity of the expected Bragg reflections, as well as satellite peaks on either side of the Bragg spots along \textit{k$_y$}. These unexpected features are true for every \(l\) = odd value (i.e., \(l\) = 1, 3, 5, etc.). Our simulated diffraction patterns for an expanded $7 \times 7$ cell (Fig.~\ref{Fig:Structure}e-f) replicate the observed precession images for the \(hk3\) and \(2kl\) planes (\(14kl\) in the expanded cell), as shown in (c) and (f), respectively. The inset in (f) provides a magnified view at \(l\) = 3, where the satellites along \textit{k$_y$} are observed.}
\end{figure*}

Careful examination of both the x-ray and neutron diffraction data revealed that the discrepancies with the HfFe$_6$Ge$_6$-type model arise from very weak intensity at some families of Bragg reflections and the presence of additional satellite reflections around them. Notably, it was at the instrument XtremeD at ILL that we first observed these satellite reflections (see Appendix~\ref{Appen:supercell_XtremeD}), motivating further x-ray diffraction experiments. The satellite reflections are particularly evident in the reconstruction of reciprocal space (calculated precession images) centered on planes \(hkl\), with \(l\) = 3 and \(h\) = 2, from x-ray diffraction data (Fig.~\ref{Fig:Precession_hk3-2kl}b,e). On the \(hk3\) plane (Fig.~\ref{Fig:Precession_hk3-2kl}b), we clearly observe groups of satellite reflections around the expected Bragg reflections (e.g. 003, 103, and 013 marked in blue in Fig.~\ref{Fig:Precession_hk3-2kl}a), as well as along the $h$ and $k$ directions, forming a triangular grid. Meanwhile, on the \(2kl\) plane (Fig.~\ref{Fig:Precession_hk3-2kl}e), the apparent discrepancy in the expected peak intensity and the presence of satellite reflections are evident for all odd values of \(l\) (i.e., \(l\) = 1, 3, 5, etc.).

We note that due to the anomalous intensity and satellite reflections occurring at \(l\) = odd planes, a refinement using a structural model with a lattice parameter of \(c/2\) relative to the HfFe$_6$Ge$_6$-type cell yields an improved $wR2$ of 10.05\% (details in the Appendix~\ref{Appen:Refinement_YCo6Ge6-type}). This structural model corresponds to a YCo$_6$Ge$_6$-type cell, which is related to the HfFe$_6$Ge$_6$-type structure by a \(c/2\) translation, a halving of the unit cell along \(c\) and a resulting 50\% occupancy at the U and Sn atomic positions along the \(c\) axis (positions \(1a\) and \(2c\) in Table~\ref{Table:Ref_YCo6Ge6-type} of Appendix~\ref{Appen:Refinement_YCo6Ge6-type}, Fig.~\ref{Fig:YCo6Ge6-type}). This cell type is therefore often referred to as a disordered Y$_{0.5}$Co$_3$Ge$_3$-type structure. During this study, a monoclinic distortion of the YCo$_6$Ge$_6$-type cell with an incommensurate structural wave vector was independently reported for UV$_6$Sn$_6$ with even better $wR2$~\cite{Thomas2025UV6Sn6}. The halving of the unit cell along \(c\) means that the \(l\) = odd planes are no longer integers, so Bragg reflections are not expected there, requiring an additional incommensurate structural wave vector. The YCo$_6$Ge$_6$-type structure has also been reported for the compound UFe$_6$Ge$_6$~\cite{Goncalves1994JAC}, and was originally reported for DyFe$_6$Sn$_6$ and HoFe$_6$Sn$_6$~\cite{koretskaya_new_1986}. However, similar to the latter cases, we found that this YCo$_6$Ge$_6$-type model does not accurately represent our UV$_6$Sn$_6$ structure.

We discovered that the satellite reflections observed in the diffraction patterns of UV$_6$Sn$_6$ can be accounted for by a supercell arising from a modulation of the HfFe$_6$Ge$_6$-type parent cell. The modulation corresponds to a periodic expansion of the parent unit cell by a factor of 7 along the \(a\) and \(b\) axes (Fig.~\ref{Fig:Structure}d-e), yielding a $7\times 7$ superstructure in the \(ab\) plane. There is no evidence of cell expansion along \(c\). The crystal symmetry remains hexagonal, \(P6/mmm\), and the supercell lattice parameters are \(a = b = 38.548(2)\) Å; \(c = 9.154(1)\) Å, with \(\alpha = \beta = 90^\circ; \gamma = 120^\circ\). As detailed in the Appendix~\ref{Appen:RT6X6_supercells}, this modulation arises from the ordering of hexagonal cavities occupied by uranium in UV$_6$Sn$_6$ (Fig.~\ref{Fig:Structure}c,e,f; these cavities are also referred to as \textit{voids} in the literature). This type of ordering is observed in other \textit{RT}$_6$\textit{X}$_6$ compounds, leading to patterns that extend the structures into supercells with modulations ranging from 2 to 68 parent unit cells~\cite{Venturini2008CMA,Fredrickson2008JACS}. These atomic arrangements are explained by electrostatic and steric (size-related) effects, which impose preferences regarding the simultaneous occupation of neighboring cavities along \(c\)~\cite{Fredrickson2008JACS}. We note that in contrast to other modulated structures such as $R$Fe$_6$Sn$_6$ with $R$ = Er, Y, Tb~\cite{Fredrickson2008JACS}, the modulation in our UV$_6$Sn$_6$ superstructure retains the hexagonal symmetry. An orthorhombic or monoclinic cell with a single $q$ modulation such as the one proposed in Ref.~\cite{Thomas2025UV6Sn6} only accounts for a pair of satellite reflections rather than the six we observe in the $hk3$ plane in Fig.~\ref{Fig:Precession_hk3-2kl}b. It is possible that our observations result from triply twinned crystals with the monoclinic structure proposed by Thomas \textit{et al}~\cite{Thomas2025UV6Sn6}. From diffraction data collected from various single crystals of different sizes--including some of dimensions comparable to those reported for x-ray diffraction in Ref.~\cite{Thomas2025UV6Sn6}, we consistently observe satellite reflections that retain the global hexagonal symmetry within the $ab$ plane. This is also observed in the neutron diffraction data shown in Fig.~\ref{Fig:Raw_XtremeD} of Appendix~\ref{Appen:supercell_XtremeD}. It is thus likely that we have crystallized a high-temperature structure with a hexagonal symmetry, different from the monoclinic structure reported in Ref.~\cite{Thomas2025UV6Sn6}. This is possible considering the different temperature profile compared to that in Ref.~\cite{Thomas2025UV6Sn6}. A sensitivity to the thermal treatment for the stabilization of different crystal structures has been reported for other $RT_6X_6$ phases~\cite{el_idrissi_crystal_1991,malaman_magnetic_1997}, and may be critical for understanding the tunability of the magnetic and charge ordered phases of $R$Mn$_6$Sn$_6$, $R$Fe$_6$Ge$_6$ and $Ln$Nb$_6$Sn$_6$ ($Ln$ = lanthanide), respectively~\cite{Wang2023NRP,riberolles_new_2024,Ortiz2025JACS}.

In our proposed extended cell of UV$_6$Sn$_6$, the uranium-occupied hexagonal cavities shift along the \(c\) axis by \(c/2\) from their original position in the HfFe$_6$Ge$_6$-type parent structure, as illustrated in Fig.~\ref{Fig:Structure}e. Labelling the shifts of the hexagonal cavities as $z=0$ for the original position and $z=$~½ for the \(c/2\) shifted (in orange and blue, respectively, in Fig.~\ref{Fig:Structure}e), the resulting pattern along the \(a\) and \(b\) crystallographic axes is 0, 0, ½, ½, ½, ½, 0. There is nearly an equal number of \(z\)~= 0 and \(z\)~= ½ sites in the supercell (25 and 24). This further explains why a refinement against the YCo$_6$Ge$_6$-type cell, with 50\% occupancy at the U sites, yields a reasonable fit. Refinement of this supercell is challenging due to the large number of parameters and data points. However, the simulated diffraction patterns for our proposed UV$_6$Sn$_6$ expanded cell accurately replicate the diffracted spots observed in the reconstructed precession images, as demonstrated in Figs.~\ref{Fig:Precession_hk3-2kl}c,f.

As noted above, the superstructure of UV$_6$Sn$_6$ is significantly different from other $RT_6X_6$ compounds. While the atomic sizes and lattice parameters in UV$_6$Sn$_6$ are similar to those of other $R$V$_6$Sn$_6$ compounds (where $R$ = Y, Sm, Gd, Dy-Lu), most of these compounds typically adopt either the undistorted, non modulated HfFe$_6$Ge$_6$-type structure or the partially disordered SmMn$_6$Ge$_6$-type structure (with the YCo$_6$Ge$_6$-type at the fully disordered extreme)~\cite{malaman_magnetic_1997,Ortiz2025JACS}. We observe that, generally, larger radii at $R$ seem to lead to modulated structures, which could be explained from the expansion of the hexagonal cages (Fig.~\ref{Fig:Structure}c), which imposes additional steric constraints. The superstructure in UV$_6$Sn$_6$ might also be driven by the spatially extended nature of the U $5f$ orbitals.

As will be discussed below, we do not find evidence of a CDW state in UV$_6$Sn$_6$. This contrasts with the closely related compound ScV$_6$Sn$_6$~\cite{Arachchige2022PRL, lee_nature_2024}. The radius of Sc, $\sim0.89$~\AA, is considerably smaller than that of U, $\sim1.14$~\AA, (VIII-coordinate Shannon radius~\cite{shannon_revised_1976} as compared in Ref.~\cite{Ortiz2025JACS}). An additional consequence of the expansion of the hexagonal cavities when occupied by atoms with increasing radii is the reduction of Sn–Sn distances. This bond shortening limits the displacement of Sn along the $c$ axis, thus suppressing the ``rattling" $R$-Sn-Sn-$R$ bond modulation, identified as a potential CDW instability in ScV$_6$Sn$_6$~\cite{lee_nature_2024, Ortiz2025JACS, riedel_magnetic_2025}.



\subsection{Magnetization}
\label{magnetization}

\begin{figure}[!htb]
\centering
\includegraphics[width=\linewidth]{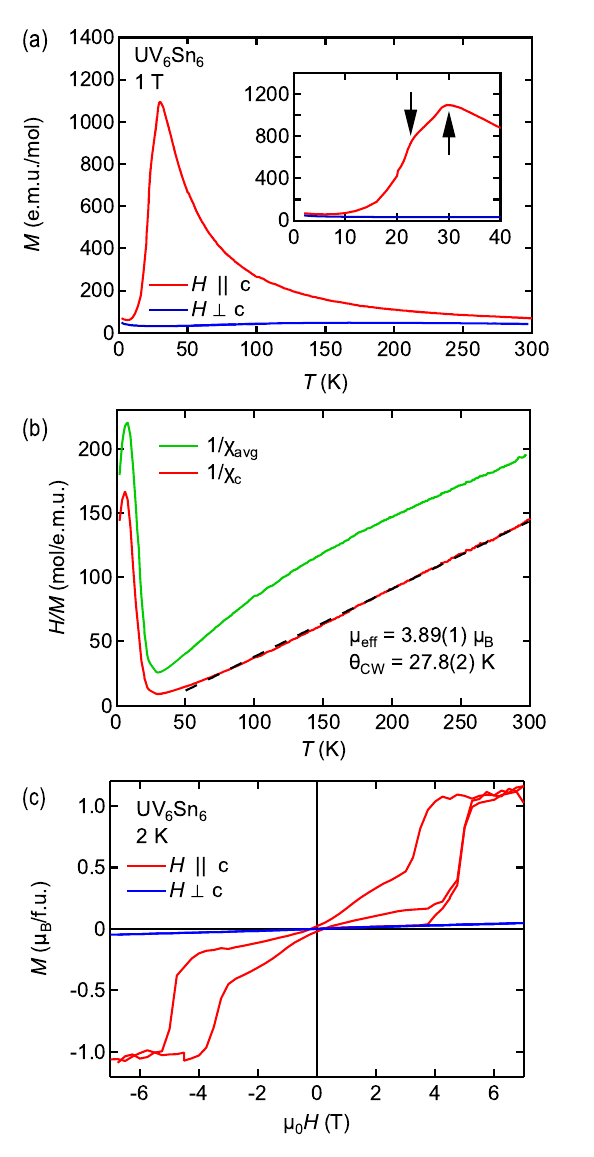}
\caption{(a) Temperature dependence of the magnetization under an applied magnetic field of 1\,T parallel and perpendicular to the $c$ axis. Inset: Enlarged view at low temperatures. The arrows indicate the anomalies at the two transitions. (b) Temperature dependence of the inverse susceptibility at 1\,T along the c-axis and for a polycrystalline average: $\chi_{avg}=\frac{1}{3}\chi_{c}+\frac{2}{3}\chi_{ab}$. The dashed line is a fit to the Curie-Weiss law. (c) Magnetic field dependence of the magnetization at 2\,K for the field parallel and perpendicular to the $c$ axis.\label{Magnab}} 
\end{figure}

The magnetization data collected from a single crystal of UV$_6$Sn$_6$, with a field of 1\,T applied parallel and perpendicular to the crystallographic $c$ axis, is presented in Fig.~\ref{Magnab}. The magnetization as a function of temperature reveals a strong magnetic anisotropy, with the $c$ axis being the easy magnetization axis (Fig.~\ref{Magnab}a). Along the $ab$-plane, the magnetization remains low and nearly constant across the entire temperature range. Notably, the magnetization parallel to the $c$ axis exhibits a clear maximum at $T_{N_1}\approx29$\,K. A second transition occurring at approximately 23\,K is also revealed as a kink in the magnetization curve (inset Fig. \ref{Magnab}a and Fig.~\ref{Fig:comparison} in Appendix \ref{comparison}). The effective magnetic moment, derived from the Curie-Weiss fit which holds down to about 70~K (Fig.~\ref{Magnab}b) along the $c$ axis, is found to be $3.89(1)$\,$\mu_B$/U. This value is slightly higher than the effective moment of the U$^{3+}$ or U$^{4+}$ configurations. The polycrystalline averaged susceptibility ($\chi_{avg}=\frac{1}{3}\chi_{c}+\frac{2}{3}\chi_{ab}$) does not follow the Curie-Weiss behavior over a large temperature range, indicating the importance of the crystal electric field (CEF). A detailed CEF analysis was recently reported~\cite{Thomas2025UV6Sn6}. The results indicate that the first excited CEF level is between 18 to 25\,meV, consistent with an isolated ground state of easy-axis moments~\cite{Thomas2025UV6Sn6}. 

The magnetization  isotherms $M(H)$ collected at 2\,K are shown in Fig.~\ref{Magnab}c. When the magnetic field is applied parallel to the $c$ axis, a clear hysteresis effect is observed, along with two notable changes in the magnetization. There is a sharp change in slope at approximately 3 T, followed by a tendency towards saturation starting around 4 T. The magnetization at 6~T is roughly 1~$\mu_B$, which is lower than the ordered moment of the free U ion. However, as discussed below, the magnetization at 6\,T does not correspond to the fully polarized moment.

\subsection{Resistivity}

\begin{figure}[!htb]
\centering
\includegraphics[width=\linewidth]{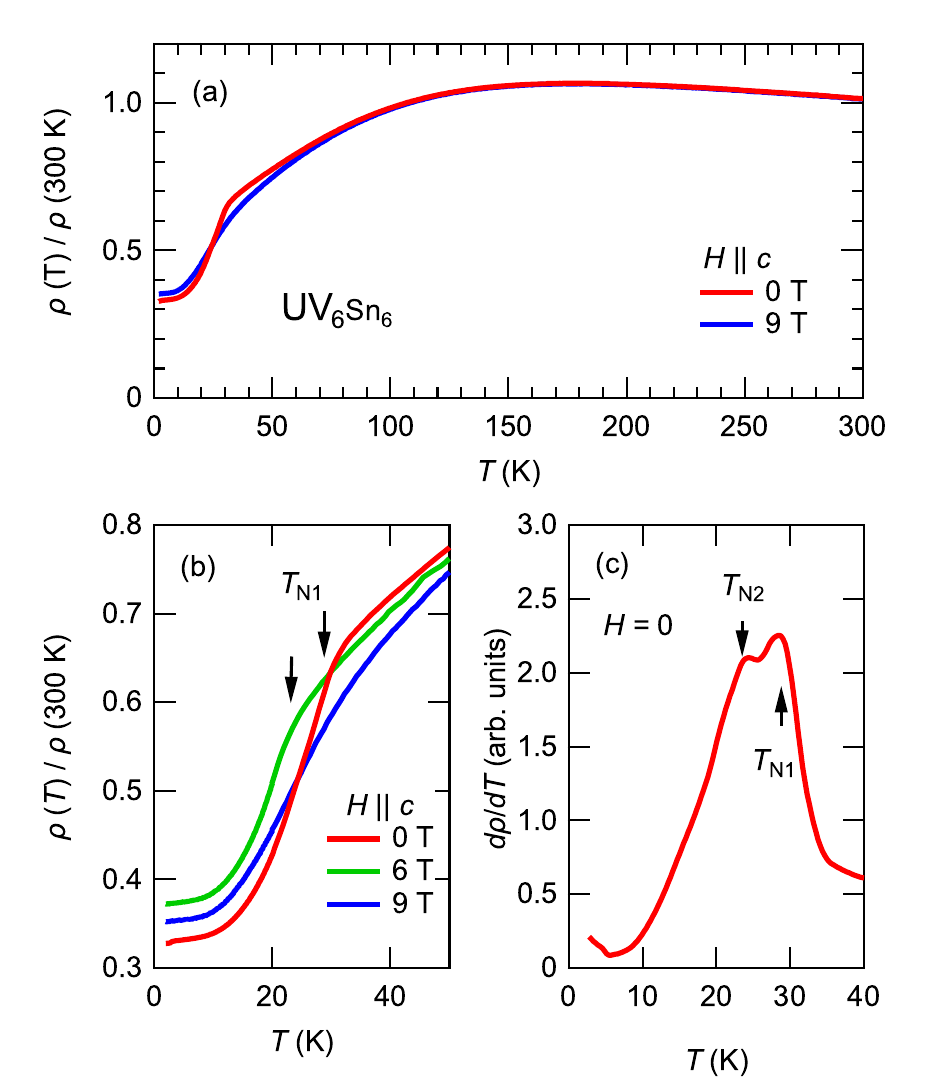}
\caption{\label{resistivity}(a) Temperature dependence of the electrical resistivity at magnetic fields of 0\,T and at 9\,T applied along the $c$ axis,  normalized to the value at 300~K. (b) Zoom of the low temperature range for fields of 0, 6, and 9~T. The arrows indicate the magnetic ordering temperature $T_{N1}$. (c) Temperature derivative of the resistivity at zero field as a function of temperature. The arrows show the magnetic transition temperatures $T_{N1}$ and $T_{N2}$.} 
\end{figure}

The temperature dependence of the normalized electric resistivity at magnetic fields of 0 and 9\,T applied along the $c$ axis is presented in Fig.~\ref{resistivity}. The resistivity ratio $RRR = R (300 {\rm K})/R (1.8 {\rm \,K}) \approx 3$ is rather low. A similar value was reported recently~\cite{Thomas2025UV6Sn6}. On cooling the resistivity first increases slightly and shows a maximum around 180~K.  At high temperatures, no magnetoresistance occurs and down to 90~K the temperature dependence at 9~T is similar to that at zero magnetic field. On further cooling, the resistivity shows a marked kink at the onset of the magnetic order near $T_{N1} \simeq 29~K$ (see Fig.~\ref{resistivity}b). At 6\,T, the magnetic transition is still visible by a clear kink, while it disappears at 9~T.

Figure \ref{resistivity}c presents the derivative of the normalized resistivity as a function of temperature. It shows a  mean-field-like anomaly at the transition, as expected from the Fisher-Langer behavior $d\rho/dT \propto C_P$, often observed in metallic systems \cite{Fisher1968PRL, Knebel2001PRB}. In particular, we do not observe an increase of resistivity upon lowering the temperature through the magnetic transition, unlike other incommensurate antiferromagnets due to the opening of a superzone gap in the electronic band structure~\cite{Mackintosh1962PRL,Elliott1963PPS,Becker1997PRL,Onimaru2008JPSJ,Feng2013PNAS,Mishra2025PRM}. This may be due to the electrical current direction being perpendicular to the magnetic wave vector (see neutron scattering and the magnetic structure discussion below). 

In the temperature derivative of the resistivity we observe a second anomaly at $T_{N2} \simeq 23$~K. This indicates a change of the magnetic structure, as discussed below in section \ref{neutron}.

\subsection{Heat capacity}

\begin{figure}[!htb]
\centering
\includegraphics[width=\linewidth]{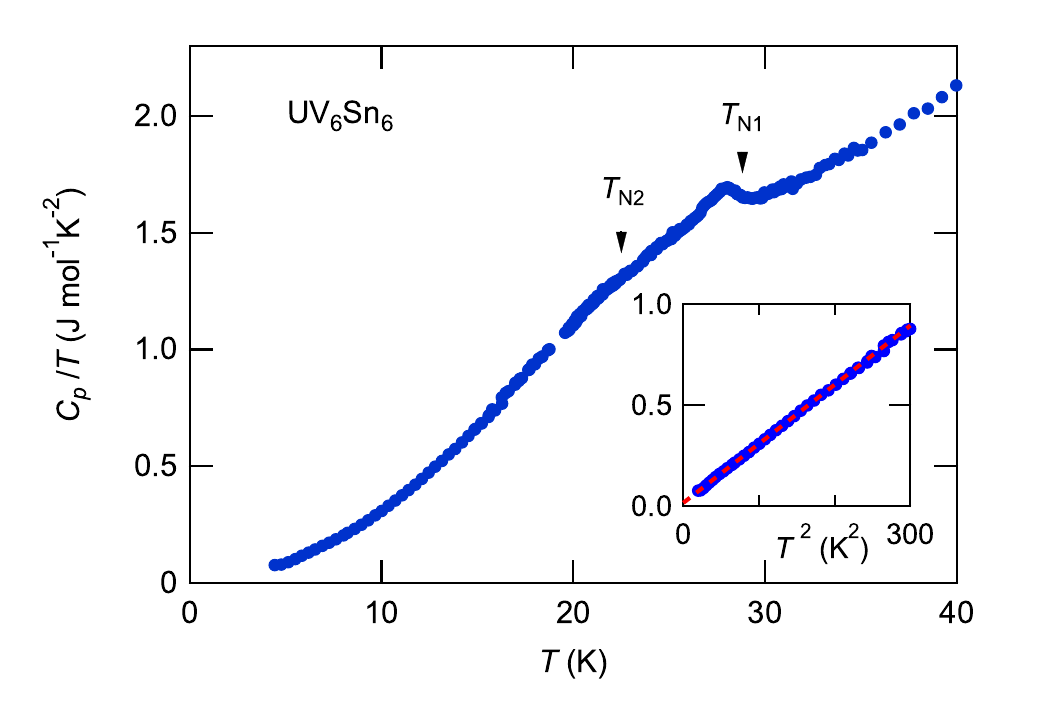}
\caption{\label{Cp_T}Specific heat divided by temperature vs temperature of UV$_6$Sn$_6$ at zero magnetic field. The arrows indicate the position of $T_{N1} = 28.8$~K and $T_{N2} = 23$~K. The inset shows $C_p$ as a function of $T^2$. The dashed line is a linear fit to $C_p/T =\gamma + \beta T^2$ and we determine $\gamma = 15~$mJmol$^{-1}$K$^{-2}$ and $\beta = 2.95$~mJmol$^{-1}$K$^{-4}$.} 
\end{figure}

\begin{figure}[!htb]
\centering
\includegraphics[width=\linewidth]{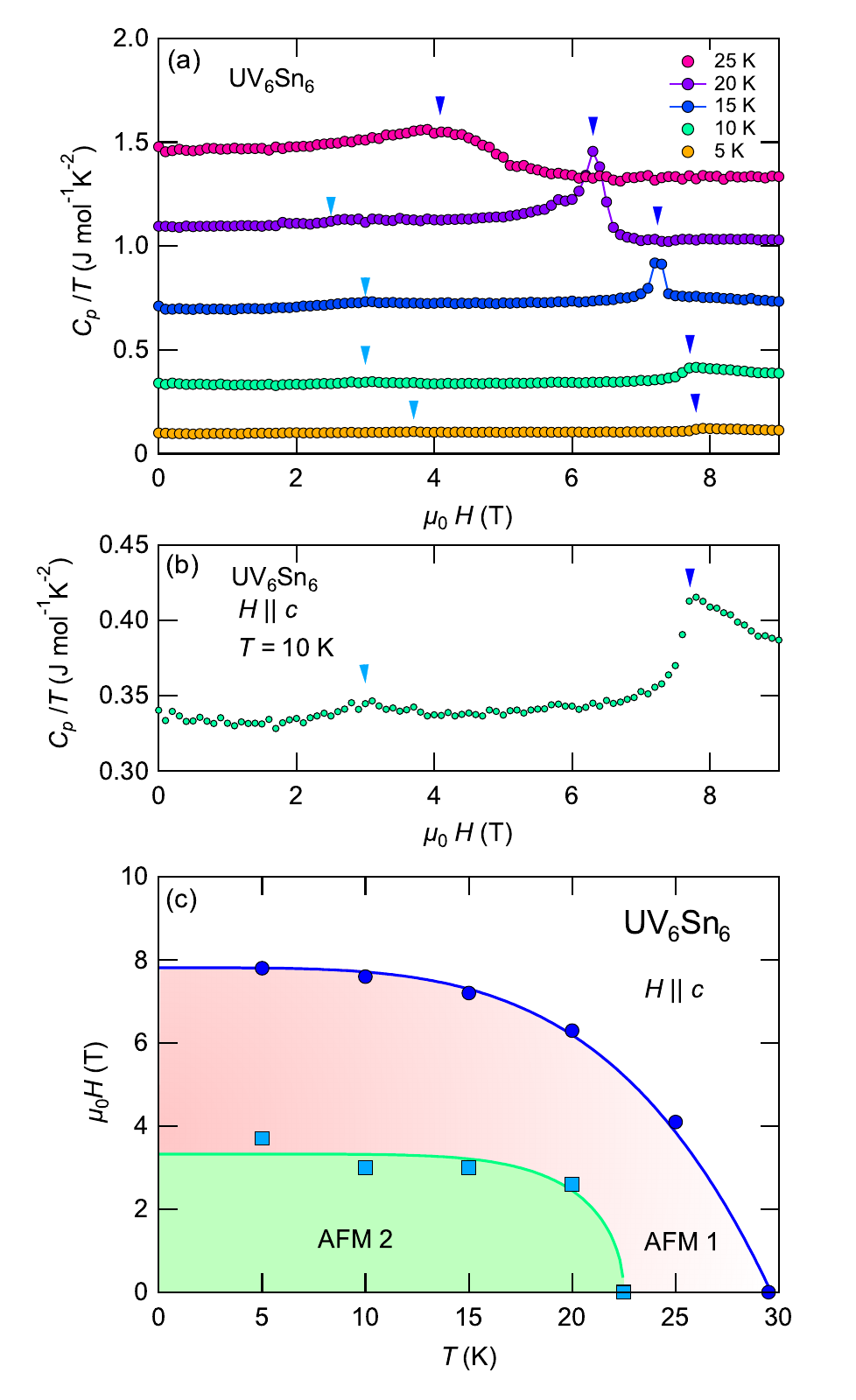}
\caption{\label{Cp_H}(a) Specific heat divided by temperature at selected \textit{T} values as a function of increasing magnetic fields. (b) Field dependence of the specific heat at 10~K in an enlarged view. (c) Magnetic phase diagram for field $H \parallel c$ from the field dependent specific heat measurements (increasing fields).  (Lines are guides to the eyes.) An additional phase boundary within the AFM1 region was inferred from thermal expansion measurements~\cite{Thomas2025UV6Sn6}.} 
\end{figure}


The temperature dependence of the specific heat divided by temperature ($C_p/T$) below 40\,K is displayed in Fig.~\ref{Cp_T}. $C_p/T$ shows two anomalies which correspond to the magnetic transitions. While the anomaly at $T_{N1} \simeq 29$~K is well defined and looks mean-field-like, the anomaly at $T_{N2} \simeq 23$~K is tiny and rather broad. This indicates that the entropy change at $T_{N2}$ is rather small and is due to a change in the magnetic structure. The inset in  Fig.~\ref{Cp_T} shows $C_p / T$ as a function of $T^2$. From the linear fit, we extract the Sommerfeld coefficient $\gamma = 15$~mJ\,mol$^{-1}$K$^{-2}$ and the coefficient of the cubic term $\beta = 2.95$~mJ\,mol$^{-1}$K$^{-4}$. This shows that UV$_6$Sn$_6$ is a weakly correlated U compound and magnetism is mainly due to localized 5$f$ electrons. Although a larger $\gamma$ was obtained, a similar conclusion was reached in a recent study~\cite{Thomas2025UV6Sn6}. This is also in agreement with the susceptibility measurements, where the large effective moment is observed almost down to $T_{N1}$ (see Fig.~\ref{Magnab}b). It is also in accordance with the large U-U spacing in the unit cell, which is far above the Hill limit~\cite{Hill1970}.

A combined plot of $dM/dT$, $C_p / T$, and $d\rho/dT$ as a function of $T$ is presented in Fig.~\ref{Fig:comparison} in Appendix \ref{comparison}. $T_{N1}$ and $T_{N2}$ are confirmed from the three data sets.

In Fig.~\ref{Cp_H}a, we show the field dependence of the specific heat at various temperatures for increasing magnetic fields. As shown explicitly in Fig.~\ref{Cp_H}b for the curve at 10 K, a first increase of the specific heat is observed at 2.5~K while a very marked anomaly occurs at 7.7~K. While the low field anomaly for all temperatures is very small, the high field anomaly appears as a very pronounced peak with a first-order like behavior at 15 and 20\,K. From these anomalies, a magnetic phase diagram can be proposed as illustrated in Fig.~\ref{Cp_H}c. The low field anomaly is associated with a change in the magnetic structure, as discussed below.

\subsection{Neutron scattering and magnetic structure}
\label{neutron}

\begin{figure}
\includegraphics[width=9cm]{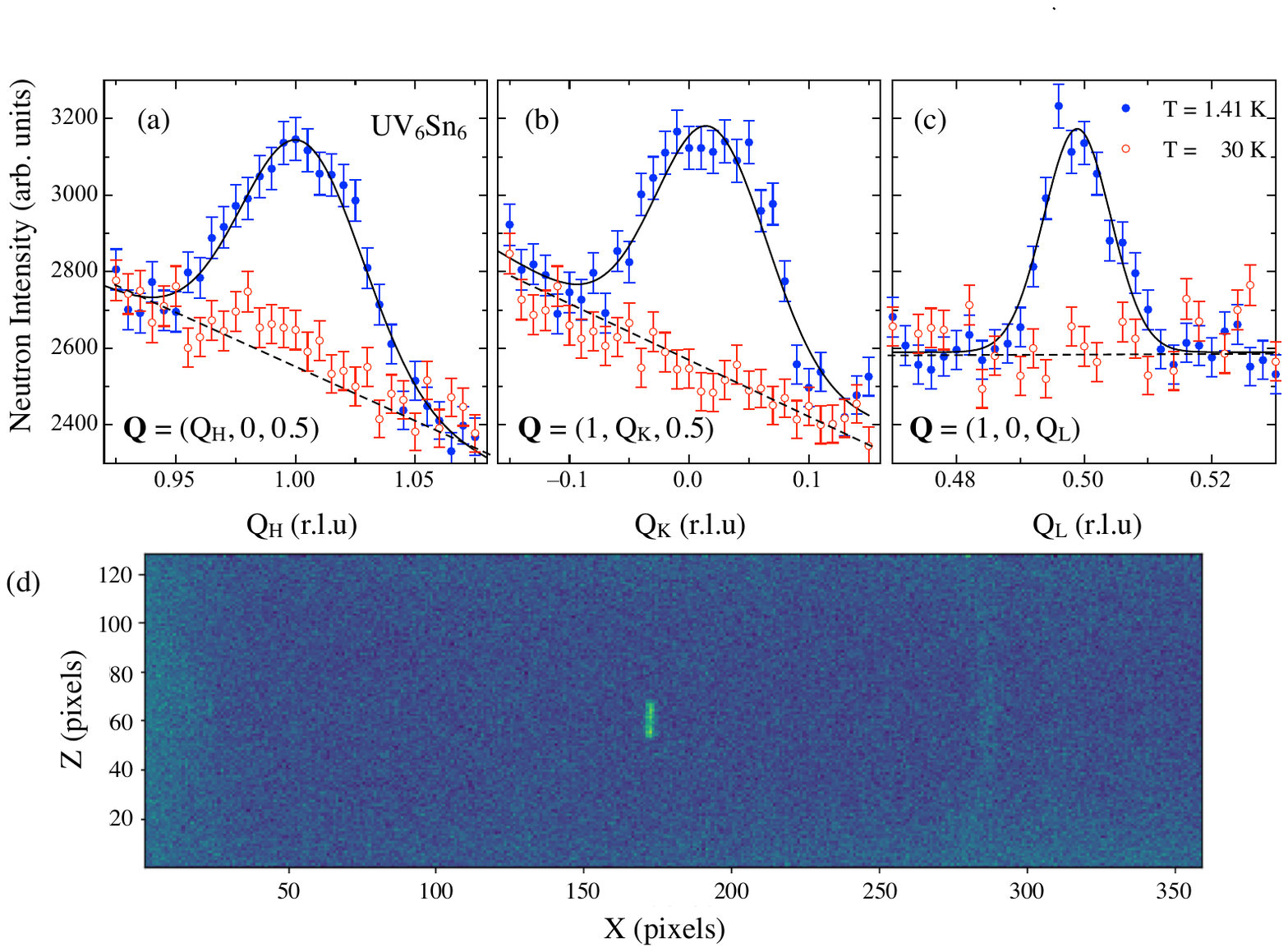}
\caption{\label{rawdata} Momentum transfer cuts obtained on D23 at 1.41 K (full blue circles) and 30 K (open red circles) along (a) ${\bf Q}=(Q_H, 0, 0.5)$, (b) along ${\bf Q}=(1, Q_K, 0.5)$, and (c) ${\bf Q}=(1, 0, Q_H)$. Full lines are fits to Gaussian and dashed lines serve as guides to the eyes. (d) Neutron intensity obtained on the XtremeD detector bank at 24.9 K for the sample rotation angle achieving diffraction condition for the peak (1, 0, 0.45) (green spot).}
\end{figure}

\begin{figure}
\includegraphics[width=9cm]{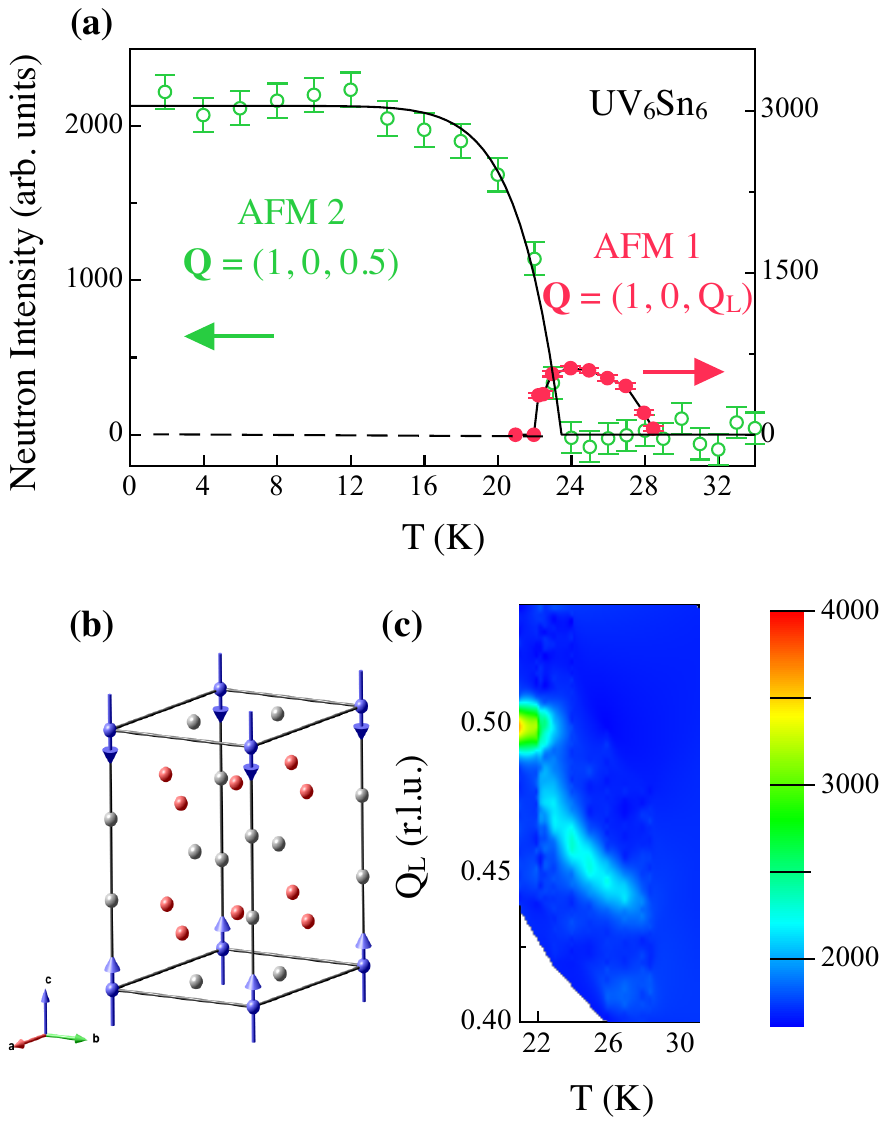}
\caption{\label{Tdep} (a) Temperature dependence of the intensity at ${\bf Q}=(1, 0, 0.5)$ (left scale) and of the integrated intensity at ${\bf Q}=(1, 0, Q_L)$  (right scale). (b) Sketch of the antiferromagnetic ground state structure (color of atoms as in Fig.~\ref{Fig:Structure}). (c) Color-coded temperature dependence of the neutron intensity as a function of temperature and $Q_L$.}
\end{figure}

Mapping of the reciprocal space along the high-symmetry lines of the Brillouin zone was carried out on D23 with the copper monochromator to identify the magnetic ordering wave vector at 1.41 K. The outcome is the identification of a single wave vector that is described below. Given the complexity of the UV$_6$Sn$_6$ crystal structure (Fig.~\ref{Fig:Structure}), we approach the magnetic structure characterization from the HfFe$_6$Ge$_6$-type parent cell.  Figure \ref{rawdata} shows scans around the momentum transfer ${\bf Q}=(1, 0 , 0.5)$ along the directions $Q_H$ (panel a), $Q_K$ (panel b) and $Q_L$ (panel c) respectively, at 1.5 K and 30 K. The coordinates of the momentum transfer $\bf{Q}$ are given in reciprocal lattice units (r.l.u.). These data clearly establish that the position of the magnetic peak is ${\bf Q}=(1, 0 , 0.5)$, which corresponds to the propagation vector $\bf{k}$=(0, 0, 0.5).  The magnetic ground state is thus antiferromagnetic (commensurate). The same magnetic wave vector was recently reported from powder neutron diffraction at 2\,K~\cite{Thomas2025UV6Sn6}. The temperature dependence of the intensity at the peak position ${\bf Q}=(1, 0 , 0.5)$ is shown in Fig.~\ref{Tdep}a. The signal vanishes at $T_{N2}$ = 23.5 K. To determine the magnetic structure, a collection of magnetic peaks was undertaken and is limited to eight peaks (7 independent ones) due to the weak signal associated with the small sample for neutron diffraction experiment. The scale factor and the extinction correction are obtained from  nuclear peaks (44 independent reflections). Among the structural parameters given in Table \ref{Table:Ref_HfFe6Ge6-type}, the isotropic thermal displacements (except for vanadium which has a very small coherent neutron scattering cross section) and the $z$ position of Sn3 are refined yielding values consistent with the x-ray refinement.  The magnetic form factor $f(\bf{Q})$ is fixed to the one of U$^{3+}$ in the dipolar approximation $f(Q)$=$j_0$($Q$)+$C_2$$j_2$($Q$) using the value $C_2$=1.75, where $j_0(Q)$ and $j_2(Q)$ are spherical Bessel functions \cite{Freeman1976PRB}.  All the magnetic reflections with ${\bf Q}$ parallel to the $c$ axis, namely, (0, 0, 0.5), (0, 0, 1.5) and (0, 0, 2.5), have zero intensity, which, according to neutron scattering selection rules point to the fact that the magnetic moment must lie along the $c$ axis. The magnitude of the moment is calculated to be 1.43(3) $\mu_B$. The refined magnetic structure is shown in Fig.\ref{Tdep}b.

Above $T_{N2}$, the search for the ordering wave vector was performed on XtremeD at ILL. A representative acquisition on the detector bank is shown in Fig.\ref{rawdata}d for a temperature of 24.9 K. It is obtained by rotating the single crystal sample with a step of 0.2$^{\circ}$ and an acquisition time of approximatively 10 minutes. A peak is observed at a position corresponding to the momentum transfer ${\bf Q}=(1, 0, 0.45)$. This incommensurate position is found to be temperature dependent and further measurements were performed on D23 to investigate this using the PG monochromator. The obtained temperature variation of the incommensurate position is shown in Fig.\ref{Tdep}c through the temperature dependence of the neutron intensity as a function of $Q_L$. Hence, the AFM1 phase is incommensurate at zero magnetic field (the evolution of the propagation vector under magnetic field where additional phase boundaries are observed~\cite{Thomas2025UV6Sn6} is not addressed in the present paper). The temperature dependence of the $Q_L$-direction integrated intensity is shown in Fig.~\ref{Tdep}a.  The intensity collapses at $T_{N1}\approx29$\,K. There is a temperature range near $T_{N2}$, where both antiferromagnetic and incommensurate phases coexist, in agreement with a transition of the first order. In addition, the mapping capabilities of XtremeD allow us to confirm that only the antiferromagnetic propagation vector ${\bf Q}=(1, 0 , 0.5)$ is present at 1.5\,K.

Among rare-earth based RV$_{6}$Sn$_{6}$ compounds, the magnetic structures were determined for $R$ = Gd by resonant x-ray diffraction \cite{Porter2023PRB} and for $R$ = Tb, Dy, Ho, Er by neutron diffraction \cite{Zhou2024PRR}. In GdV$_{6}$Sn$_{6}$, two transitions are observed at 3.8 K and 5.2 K. In the high temperature phase, only an incommensurate propagation vector, ${\bf k}$=(0, 0, $q_{IC}$), is observed, with a small temperature dependence of $q_{IC}$ ranging between 0.46 and 0.47. Below the second transition, it coexists with a commensurate component ${\bf k}$=(0, 0, 0.5) down to the lowest measured temperature of 2 K. A lock-in at lower temperatures is suggested. The magnetic moments lie presumably predominantly in the basal plane of the structure. TbV$_{6}$Sn$_{6}$, DyV$_{6}$Sn$_{6}$ and HoV$_{6}$Sn$_{6}$ are ferromagnetic materials with transition temperatures at 4.3 K, 3 K and 2.4 K respectively and moments along the $c$ axis, except for $R$ = Dy, where a 20$^\circ$ tilt toward the $a$ axis is found. ErV$_{6}$Sn$_{6}$ is antiferromagnetic below 0.6 K with the propagation vector ${\bf k}$=(0, 0, 0.5) and the magnetic moments along the $a$ axis in the hexagonal plane. In these heavy rare-earth compounds, the magnetic moment lies in the range 6-9 $\mu_B$ (while it is not determined for GdV$_{6}$Sn$_{6}$).

The magnetic structure of UV$_{6}$Sn$_{6}$ is distinct from each one reported for its rare-earth counterparts: it adopts the antiferromagnetic propagation vector common to Gd and Er compounds but the moment direction found in the Tb and Ho ones. One can also notice the highest transition temperature by almost one order of magnitude as compared to the rare-earth systems. The magnetic moment for 1.42(3)\,$\mu_B$ is in the typical range of values found in uranium intermetallics. The ground-state antiferromagnetic phase is made of ferromagnetic planes, with presumably a strong exchange interaction given the positive Curie-Weiss temperature and with an antiferromagnetic stacking along the $c$ axis involving likely a smaller interaction. The observed full lock-in transition from the incommensurate to the antiferromagnetic structure with decreasing temperature could be related to the highest transition temperature as compared to the Gd system, where the lock-in is partial. An interesting and open question concerns the arrangement of the spins in the supercell, of which determination goes beyond the present paper. One can mention the simplest structure where all spins in the $z$ = 0 and $z$ = 1/2 layers (see notations above) of the supercell would be ferromagnetic. In this case, the global structure can be seen as composed of antiferromagnetic blocks made of the magnetic structure as described here in the HfFe$_6$Ge$_6$ cell with "sliding defects" in the structure at the boundary between $z$ = 0 and $z$ = 1/2 layers. Another related global description of the magnetic structure can be made in the YCo$_{6}$Ge$_{6}$ model, having a halving of the $c$ axis parameter and half occupancy of uranium sites (see above), where the magnetic arrangement is of type "+ + - -" with a propagation vector ${\bf k}$ = (0, 0, 0.25).

\section{Conclusion}

Our study reveals that the UV$_6$Sn$_6$ structure deviates from the HfFe$_6$Ge$_6$-type structure commonly reported for \(RT_6X_6\) systems. The structure is best described as a $7\times 7\times 1$  modulation of the HfFe$_6$Ge$_6$-type parent cell on the \(ab\) plane, driven by an ordering pattern of the uranium atoms. This model replicates the observed features in x-ray and neutron diffraction data.

Magnetic characterization from magnetization and neutron diffraction measurements demonstrate two successive magnetic transitions at $T_{N1}\approx29$\,K and $T_{N2}$~=~23.5\,K. The first transition corresponds to an incommensurate antiferromagnetic ordering that locks to a commensurate antiferromagnetic ground state at $T_{N2}$. This behavior suggests competing exchange interactions. The slightly large effective moment ($3.89(1)$\,$\mu_B$/U) and strong uniaxial anisotropy along the $c$ axis point to significant spin-orbit coupling, a hallmark of uranium-based systems.

Our paper opens several future research avenues. The superstructure of UV$_6$Sn$_6$ emphasizes the unique role of uranium’s $5f$ orbitals. The complexity of the crystallographic and magnetic structures invites further investigation. Exploring pressure or chemical doping could help clarify the interplay between uranium’s $5f$-driven structural modulations and the vanadium kagome network in UV$_6$Sn$_6$. Applying pressure may suppress the $7\times 7\times 1$ supercell, providing insights into the relationship between steric and electronic contributions to the modulation, and shed some light on competing orders. Doping may tune the hybridization between the $5f$ and $3d$ orbitals in the kagome network. Such studies would advance our understanding of both the structural complexity of UV$_6$Sn$_6$ and the broader class of 166 kagome compounds.

\subsection{Acknowledgements}

We thank E. Bauer, P. Rosa, S. Thomas for letting us know about Ref.~\cite{Thomas2025UV6Sn6} as well as for sharing additional x-ray diffraction data. M.A.P. thanks F. Denis Romero for fruitful discussions about crystallography. M.A.P. and V.T. acknowledge funding from the Laboratoire d’excellence LANEF in Grenoble (ANR-10-LABX-51-01) and the support from the US National Science Foundation (NSF) Grant No. 2201516 under the Accelnet program of the Office of International Science and Engineering (OISE). P.L.B. was supported by QuantEdu-France (ANR-22-CMAS-0001). We further acknowledge financial support from the French National Agency for Research ANR within  Project FRESCO (ANR-20-CE30-0020) and SCATE (ANR-22-CE30-0040).

\subsection{Data availability}

The data that support the findings of this article are openly available~\cite{D23,XtremeD}. The rest of the data are available upon reasonable request from the authors.

\bibliography{biblio,Midori_2}

\appendix
\section{Appendix A: Details of structural refinement of UV$_6$Sn$_6$ using a HfFe$_6$Ge$_6$-type model}
\label{Appen:Details_HfFe6Ge6-type}

The structural parameters and anisotropic parameters assuming a HfFe$_6$Ge$_6$-type structure are presented in tables~\ref{Table:Ref_HfFe6Ge6-type} and \ref{Table:Uaniso_HfFe6Ge6-type}.

\begin{table}[H]
\centering
\caption{\label{Table:Ref_HfFe6Ge6-type} Structural parameters for UV$_6$Sn$_6$ refined against single-crystal x-ray diffraction data (Bruker D8 Venture, Mo~\textit{K}$_{\alpha}$ radiation, collected at 100 K). The isotropic displacement parameters ($U_{iso}$) are given in \AA$^2\times10^3$. The anisotropic parameters are provided in Table.~\ref{Table:Uaniso_HfFe6Ge6-type}.}
\begin{tabular}{|c|c|c|c|c|c|c|}
\hline
\textbf{Atom} & \textbf{Site} & \textbf{\(x\)} & \textbf{\(y\)} & \textbf{\(z\)} & \textbf{Occ.} & \textbf{$U_{iso}$} \\ \hline
U & \(1a\) & 0 & 0 & 0 & 0.991(5) & 9.4(3) \\ \hline
Sn1 & \(2d\) & 1/3 & 2/3 & 1/2 & 0.990(9) & 5.9(2) \\ \hline
Sn2 & \(2c\) & 1/3 & 2/3 & 0 & 0.981(9) & 5.5(2) \\ \hline
Sn3 & \(2e\) & 0 & 0 & 0.3434(1) & 1.000 & 6.2(1) \\ \hline
V & \(6i\) & 1/2 & 0 & 0.2489(3) & 1.000 & 6.9(4) \\ \hline
\multicolumn{7}{|c|}{Space group \( P6/mmm \) (No. 191)} \\
\multicolumn{7}{|c|}{$a = b = 5.5082(1)$ \AA ; $c = 9.1537(5)$ \AA ;} \\
\multicolumn{7}{|c|}{$\alpha = \beta = 90^\circ$; $\gamma = 120^\circ$} \\ \hline
\multicolumn{7}{|c|}{GOF = 1.76, \textit{wR2} = 20.76\%} \\ \hline
\end{tabular}
\end{table}

\begin{table}[H]
\centering
\caption{\label{Table:Uaniso_HfFe6Ge6-type} Anisotropic displacement parameters (\AA$^2\times10^3$) for the UV$_6$Sn$_6$ structure summarized in Table~\ref{Table:Ref_HfFe6Ge6-type}, refined using a HfFe$_6$Ge$_6$-type model against single-crystal x-ray diffraction data (Bruker D8 Venture, Mo \textit{K}$_{\alpha}$ radiation, collected at 100 K).}
\begin{tabular}{|l|l|l|l|l|l|l|}
\hline
{Atom} & \textbf{$U_{11}$} & \textbf{$U_{22}$} & \textbf{$U_{33}$} & \textbf{$U_{12}$} & \textbf{$U_{13}$} & \textbf{$U_{23}$}\\ \hline
U & 6.4(4) & 6.4(4) & 15.3(5) & 3.2(2) & 0.00 & 0.00 \\ \hline
Sn1 & 2.5(4) & 2.5(4) & 12.7(2) & 1.3(1) & 0.00 & 0.00\\ \hline
Sn2 & 2.4(3) & 2.4(3) & 11.8(2) & 1.2(1) & 0.00 & 0.00\\ \hline
Sn3 & 2.7(2) & 2.7(2) & 13.2(3) & 1.4(1) & 0.00 & 0.00\\ \hline
V & 4.2(8) & 3.9(7) & 12.5(4) & 2.0(5) & 5.0(9) & 0.00\\ \hline
\end{tabular}
\end{table}









\section{Appendix B: Evidence of superstructure in neutron diffraction}
\label{Appen:supercell_XtremeD}

\begin{figure}[!htb]
\centering
\includegraphics[width=\linewidth]{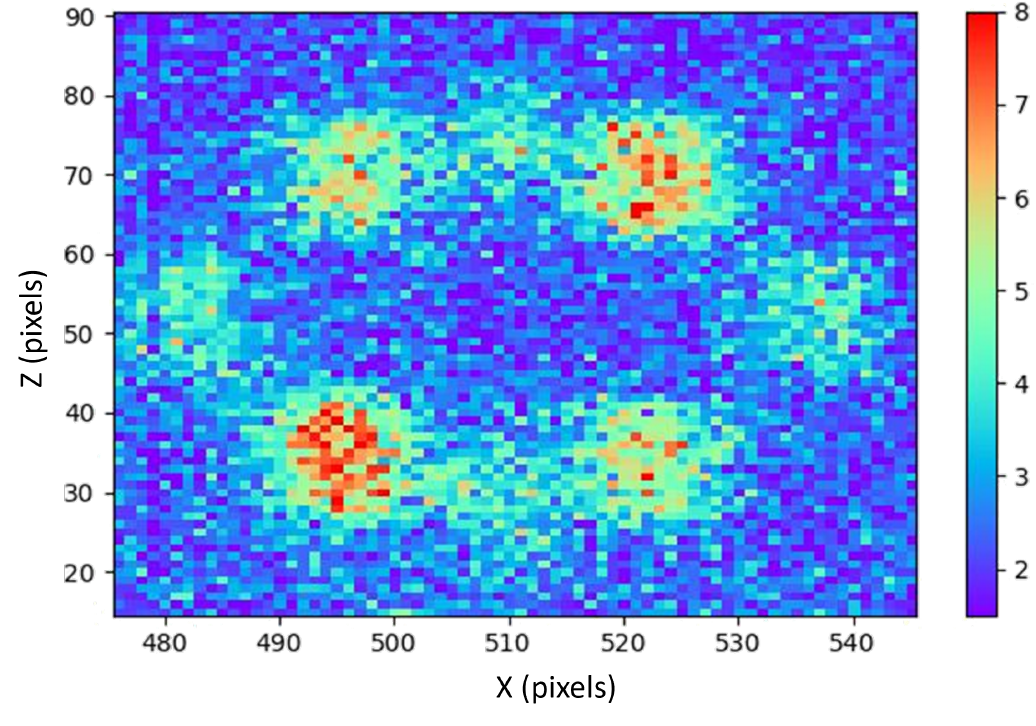}
\caption{\label{Fig:Raw_XtremeD}Raw data as observed on the large solid angle XtremeD detector.}
\end{figure}

As mentioned in the main text, the superstructure peaks associated with the UV$_6$Sn$_6$ supercell were first observed using XtremD. Figure~\ref{Fig:Raw_XtremeD} shows representative data for the satellites associated with the (2, 0, -3) reflection (in the HfFe$_6$Ge$_6$-type cell) obtained from the XtremD detector bank at 27 K. These satellites were also observed on D23 with the same positions at 24 K and 150 K, which rules out their link to the magnetic ordering below 29\,K evidenced in this paper.  

\section{Appendix C: Refinement using a YCo$_6$Ge$_6$-type model}
\label{Appen:Refinement_YCo6Ge6-type}

\begin{figure}[!htb]
\centering
\includegraphics[width=\linewidth]{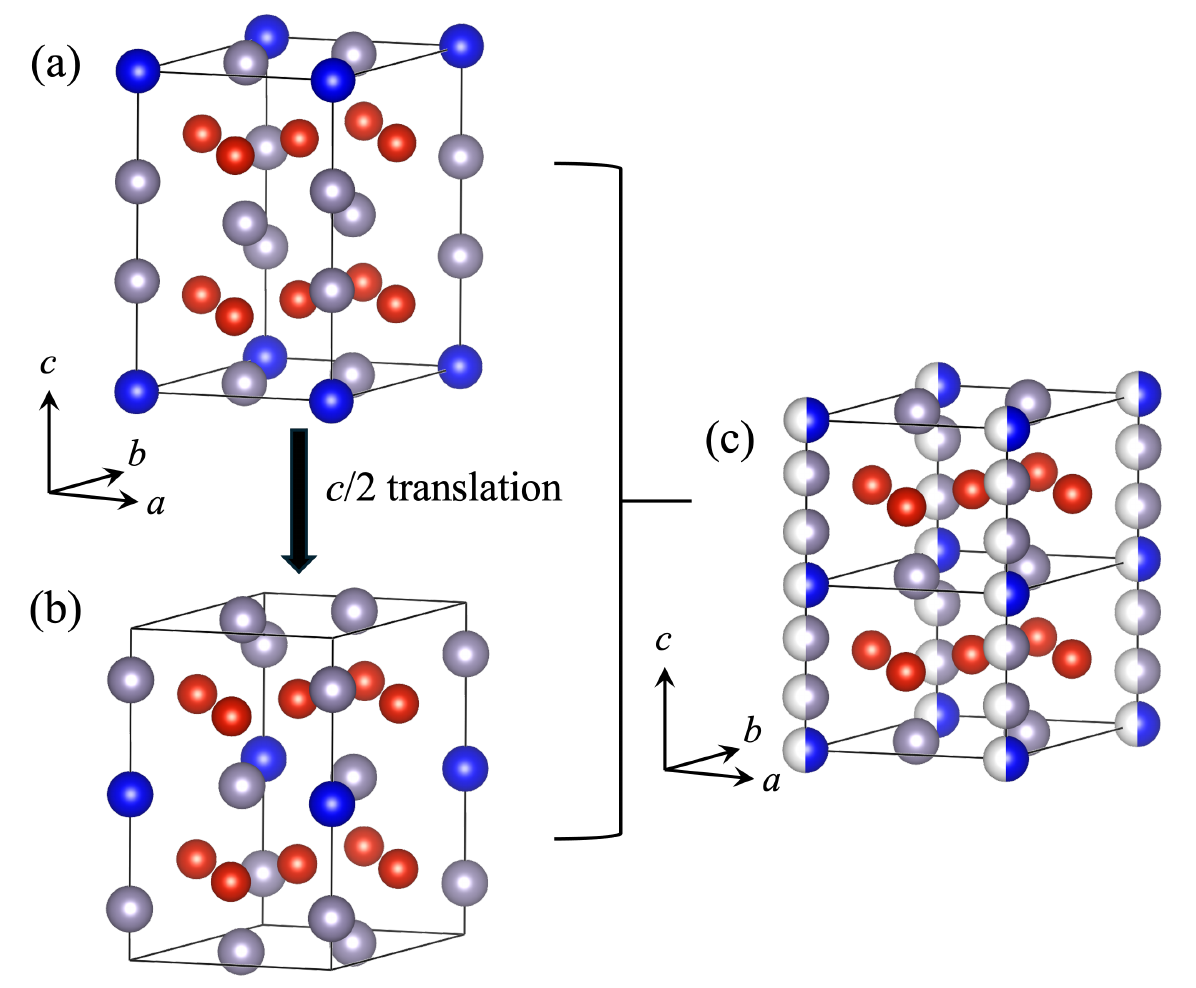}
\caption{\label{Fig:YCo6Ge6-type} The YCo$_6$Ge$_6$-type structure results from the superposition of the HfFe$_6$Ge$_6$-type cell depicted in (a) and its \(c/2\) shifted version depicted in (b), sometimes referred to as the MgFe$_6$Ge$_6$-type structure. The resulting YCo$_6$Ge$_6$-type structure shown in (c), as two stacked cells along the \(c\) axis for clarity, exhibits statistical occupancy at the crystallographic sites along the \(c\) lattice parameter (details in Table \ref{Table:Ref_YCo6Ge6-type}).}
\end{figure}

The YCo$_6$Ge$_6$-type structure is related to the HfFe$_6$Ge$_6$-type by a \(c/2\) translation, and a halving of the unit cell along the $c$ axis and of the occupancy at the U and Sn atomic positions, as illustrated in Fig. \ref{Fig:YCo6Ge6-type} (positions \(1a\) and \(2c\) in Table \ref{Table:Ref_YCo6Ge6-type}). For this model, the HfFe$_6$Ge$_6$-type cell odd-$l$ planes are no longer integers. This means that Bragg reflections are not expected at those values of $l$ during refinement. Although refinement of this YCo$_6$Ge$_6$-type model against x-ray diffraction data  leads to a reduced \(wR2\) (Table \ref{Table:Ref_YCo6Ge6-type}), it does not accurately describe the UV$_6$Sn$_6$ structure as discussed in the main text.

\begin{table}[H]
\centering
\caption{Structural parameters for UV$_6$Sn$_6$ refined using an YCo$_6$Ge$_6$-type model against single-crystal x-ray diffraction data (Rigaku XCallibur, Mo~\textit{K}$_{\alpha}$ radiation, collected at 150 K). The isotropic displacement parameters ($U_{iso}$) are given in \AA$^2\times10^3$. The anisotropic parameters are provided in Table \ref{Table:Uaniso_YCo6Ge6-type}.}
\label{Table:Ref_YCo6Ge6-type} 
\begin{tabular}{|c|c|c|c|c|c|c|}
\hline
\textbf{Atom} & \textbf{Site} & \textbf{\(x\)} & \textbf{\(y\)} & \textbf{\(z\)} & \textbf{Occ.} & \textbf{$U_{iso}$} \\ \hline
U & \(1a\) & 0 & 0 & 0 & 0.496(3) & 8.6(8) \\ \hline
Sn1 & \(2d\) & 1/3 & 2/3 & 0 & 1.000 & 7.4(7) \\ \hline
Sn2 & \(2c\) & 0 & 0 & 0.322(1) & 0.503(3) & 7.7(9) \\ \hline
V & \(6i\) & 1/2 & 0 & 1/2 & 1.000 & 7.0(9) \\ \hline
\multicolumn{7}{|c|}{Space group \( P6/mmm \) (No. 191)} \\
\multicolumn{7}{|c|}{$a = b = 5.5124(2)$ \AA ; $c = 4.5800(2)$ \AA ;} \\
\multicolumn{7}{|c|}{$\alpha = \beta = 90^\circ$; $\gamma = 120^\circ$} \\ \hline
\multicolumn{7}{|c|}{GOF = 2.41, \textit{wR2} = 10.05\%} \\ \hline
\end{tabular}
\end{table}

\begin{table}[!htb]
\caption{Anisotropic displacement parameters (\AA$^2\times10^3$) refined for the YCo$_6$Ge$_6$-type structure of UV$_6$Sn$_6$ summarized in Table~\ref{Table:Ref_YCo6Ge6-type}.}
\label{Table:Uaniso_YCo6Ge6-type} 
\begin{tabular}{|l|l|l|l|l|l|l|}
\hline
{Atom} & \textbf{$U_{11}$} & \textbf{$U_{22}$} & \textbf{$U_{33}$} & \textbf{$U_{12}$} & \textbf{$U_{13}$} & \textbf{$U_{23}$}\\ \hline
U & 6.6(8) & 6.6(8) & 12.6(8) & 3.3(7) & 0.00 & 0.00 \\ \hline
Sn1 & 5.8(7) & 5.8(7) & 10.7(7) & 2.9(3) & 0.00 & 0.00\\ \hline
Sn2 & 5.1(9) & 5.1(9) & 13.1(9) & 2.5(5) & 0.00 & 0.00\\ \hline
V & 5.1(8) & 4.5(7) & 11.4(9) & 2.2(6) & 0.2(9) & 0.00\\ \hline
\end{tabular}
\end{table}

\section{Appendix D: \(RT_6X_6\) supercells}
\label{Appen:RT6X6_supercells}

\begin{figure}[H]
\centering
\includegraphics[width=\linewidth]{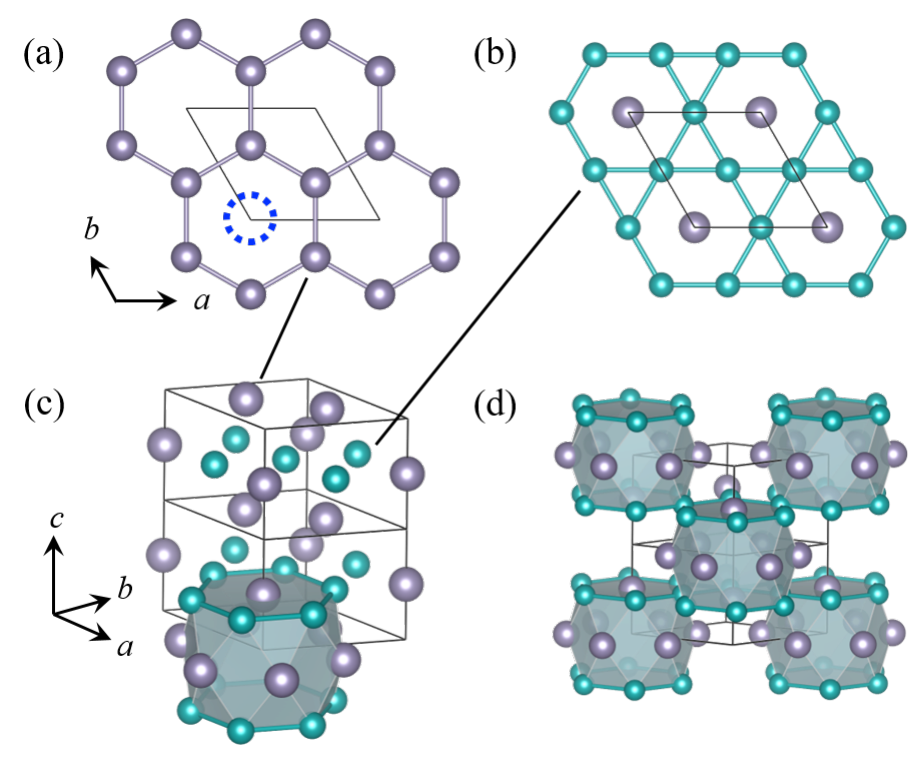}
\caption{\label{Fig:CoSn-type} The CoSn-type structure is formed by alternating layers of (a) Sn honeycomb lattices and (b) Sn-centered Co kagome networks (Sn in gray and Co in green). The resulting structure depicted in (c)--note two unit cells are stacked along \(c\)--leads to the HfFe$_6$Ge$_6$-type structure when a rare-earth is introduced in (a), at the origin of the unit cell (marked in blue). The surrounding atoms at this position form a hexagonal cavity represented as a polyhedron in (c). Electrostatic and steric (size-related) effects impose preferences regarding the simultaneous occupation of neighboring cavities along \(c\)~\cite{Fredrickson2008JACS}. These preferences may lead to patterns that extend the unit cell into a supercell such as (d).}
\end{figure}

To understand the origin of the satellite reflections observed in the diffraction patterns of UV$_6$Sn$_6$, the structure can be described as a variant of the stuffed CoSn-type cell depicted in Fig. \ref{Fig:CoSn-type}. The CoSn-type structure comprises alternating layers of Sn honeycomb lattices and Sn-centered Co kagome networks (Fig. \ref{Fig:CoSn-type}a-b). The HfFe$_6$Ge$_6$-type structure emerges when a rare-earth atom is introduced into the Sn honeycomb lattice, at the origin of the unit cell (as shown in Fig. \ref{Fig:CoSn-type}a). The atomic environment for the rare-earth atom, and for U in the case of UV$_6$Sn$_6$, can be described as a cavity with hexagonal symmetry (Fig. \ref{Fig:CoSn-type}c). The occupation of these hexagonal cavities in \(RT_6X_6\) compounds leads to patterns that extend the structures into superstructures.

\section{Appendix E: Comparison of $T_N$}
\label{comparison}

\begin{figure}[H]
\centering
\includegraphics[width=\linewidth]{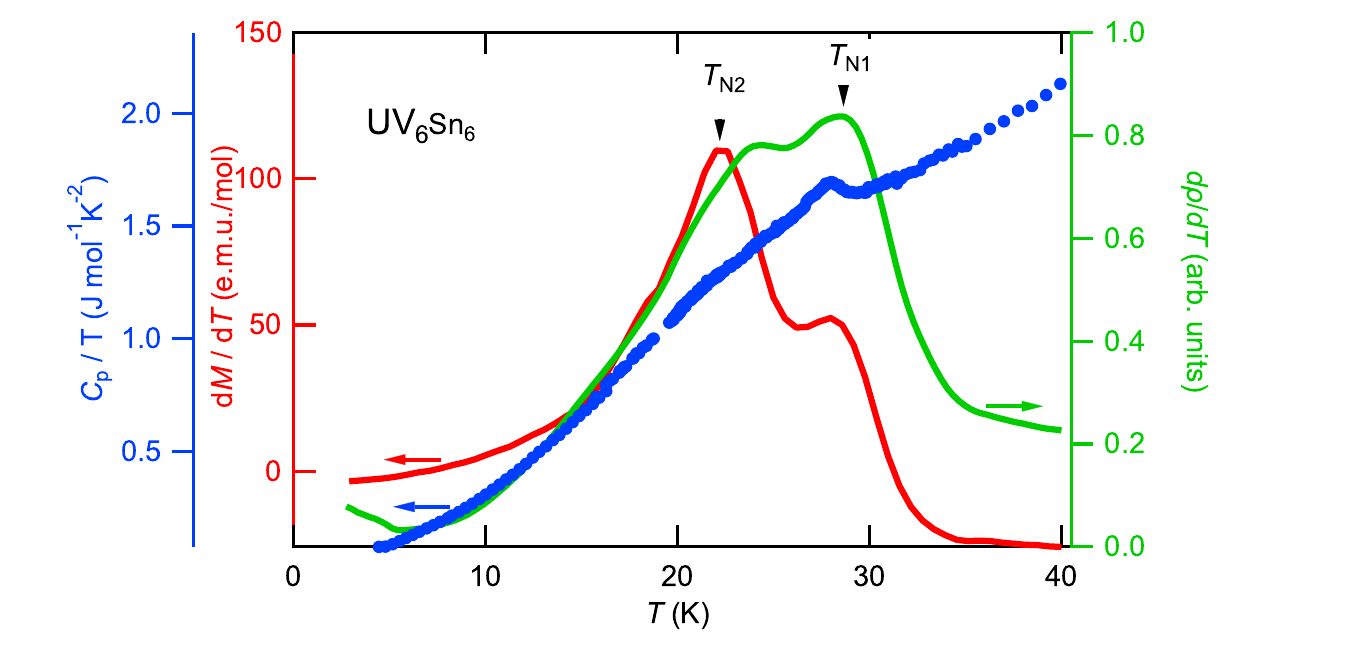}
\caption{\label{Fig:comparison} Determination of transition temperatures $T_{N1}$ and $T_{N2}$ by the temperature derivative of the magnetization for $H\parallel c$ (red) measured at 1~T, the resistivity (green, 0~T), and specific heat (blue, 0~T).}
\end{figure}

Figure~\ref{Fig:comparison}, presents a comparison of the temperature dependence of the magnetization derivative, $dM/dT$ --for a magnetic field applied along the $c$ axis, the resistivity derivative, $d\rho/dT$, and the specific heat divided by temperature, $C_p/T$. The arrows mark the magnetic transition temperatures $T_{N1}$ and $T_{N2}$. Rather good agreement is observed between the different measurements. The only slight discrepancy is that $T_{N2}$, as determined from the resistivity, appears slightly higher. This difference may arise from the use of a different single-crystal sample for that particular measurement.

\end{document}